\documentclass[prx, 10pt, twocolumn, floatfix, groupedaddress, aps, longbibliography] {revtex4-1}

\usepackage{times, mathrsfs, amsmath, amsfonts, graphics, graphicx, cancel, color, amsthm, bbm, mathtools, amssymb, physics}

\usepackage[nice]{nicefrac}
\usepackage[english]{babel}
\usepackage[margin=.75in]{geometry}
\usepackage[T1]{fontenc} 
\usepackage{capt-of}

\usepackage{xfrac,relsize,float}

\usepackage{appendix}

\usepackage[unicode=true, breaklinks=false, pdfborder={0 0 1}, backref=false, colorlinks=true, linkcolor=blue, citecolor=blue]{hyperref}

\def\bra#1{\mathinner{\langle{#1}|}}
\def\ket#1{\mathinner{|{#1}\rangle}}
\def\braket#1{\mathinner{\langle{#1}\rangle}}

\def\BraVert{\egroup\,\mid\,\bgroup}

\definecolor{Blue}{rgb}{0,0,1}
\definecolor{Red}{rgb}{1,0,0}
\definecolor{Green}{rgb}{0,1,0}
\definecolor{darkgreen}{rgb}{0,.7,0}
\definecolor{Purp}{rgb}{.2,0,.2}
\definecolor{white}{rgb}{1,1,1}

\newtheorem*{definition}{Definition}

\usepackage{url}

\def\lc{\left\lceil}   
\def\rc{\right\rceil}
\usepackage{lipsum, babel}

\begin{document}
\title{Quantum enhancement to information acquisition speed}
\author{Sebastian Horvat}
\email{sebastian.horvat@univie.ac.at}
\affiliation{University of Vienna, Faculty of Physics, Vienna Center for Quantum Science and Technology, Boltzmanngasse 5, 1090 Vienna,Austria,}
\author{Borivoje Daki\'{c}}
\email{borivoje.dakic@univie.ac.at}
\affiliation{University of Vienna, Faculty of Physics, Vienna Center for Quantum Science and Technology, Boltzmanngasse 5, 1090 Vienna,Austria,}
\affiliation{Institute for Quantum Optics and Quantum Information (IQOQI), Austrian Academy of
Sciences, Vienna, Austria}
\date{\today}

\begin{abstract}
The speed of the transmission of a physical signal from a sender to a receiver is limited by the speed of
light, regardless of the physical system being classical or quantum. In this sense, quantum mechanics can not provide any enhancement of the speed of information transmission. If instead we consider that the information needing to be transmitted is not localized at the sender’s location, but dispersed throughout space, spatial coherence might provide some enhancement. In this work, we demonstrate a quantum mechanical advantage in the speed of acquirement and transmission of information globally encoded in space. We present a task for which we prove a quadratic enhancement to the information acquisition speed using quantum information carriers with respect to their classical counterpart. Our findings can naturally be applied in situations where the information source has limited power, i.e. bounded number of signals that can be sent per unit time. 
\end{abstract}

\maketitle

\section{Introduction}
Usually, when one talks about the speed of information transmission, one envisions two parties A and B, where A aims to communicate some message to B; A then sends the message encoded in a physical system (signal or information carrier) to B and the information transmission speed is simply the speed of the signal which is limited by the speed of light. In this sense, quantum mechanics cannot provide any enhancement to the speed of information transmission. However, what if the information that needs to be transmitted is not localized at the sender's station, as it was at A in the given example? What if the information of interest is encoded in a $\textit{global}$ property of dispersed pieces of information, each localized at a different location? In this case, if we define information acquisition speed as a quantity inversely proportional to the time needed to acquire and transmit some generally global information, quantum mechanics may provide some advantage with respect to classical theory.\\
In this paper we show that preparing information carriers in spatial superposition provides an arbitrarily high speed up of an information theoretic task involving the acquisition and transmission of globally encoded information. In order to formally address the subject matter we first describe the scenario of interest and introduce the auxiliary notion of $\textit{k-way signaling behaviors}$ within a device-independent formalism \cite{device}. We then proceed by proving that a single quantum particle in spatial superposition outperforms classical particles at collecting and transmitting delocalized information. This is shown by the violation of a specific inequality which poses sharp bounds on the performance of $k$-way signalling processes. One quantum particle outperforms $N$ classical particles in a single shot (for any $N$); however, since the overhead is negligible for large $N$, we introduce a slight modification of our inequality, and show that multiple rounds enable a quadratic enhancement of the information acquisition speed (for large $N$). Our findings have a natural application in scenarios involving information sources with limited power, and are based solely on the quantum superposition principle. Our result can thus be seen in light of recent developments that put forward quantum superposition as a genuine resource for information processing, such as in two-way communication with one particle \cite{2-way}, quantum acausal processes \cite{Brukner, orders, orders2, orders3}, superpositions of directions \cite{directions}, quantum combs \cite{combs}, quantum switch \cite {switch} and quantum causal models \cite{causal models, causal models2, Causal1, Causal2}. Some of these novel phenomena have been demonstrated in recent experiments \cite{exp1, exp2, exp3, exp4, exp5, exp6}.

\section{Acquirement and transmission of delocalized information}\label{Acquirement and transmission of delocalized information}
\begin{figure}[h!]
\centering
\includegraphics[width=\columnwidth]{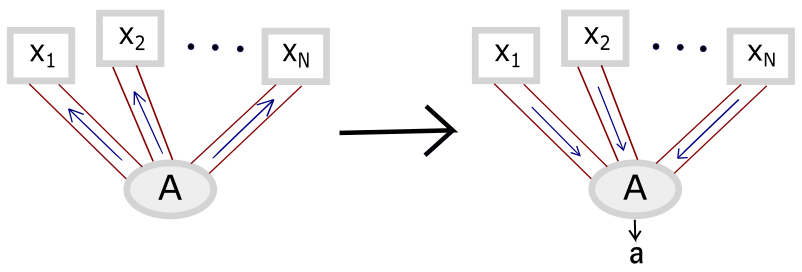}
\caption{ Alice is connected to each of the $N$ information locations via $N$ communication channels. In order to gather the required global information, Alice sends her signals towards the $N$ locations. Upon receiving back her signals, she decodes the message and produces a classical output $a$.}
\label{fig:Fig1}
\end{figure}

The scenario of interest consists of one party, whom we will refer to as Alice, and $N$ pieces of information $\left\{x_1,x_2,...,x_N \right\}$ dispersed at $N$ different locations, as pictured in Figure \ref{fig:Fig1}. Alice is connected to each of the $N$ locations with communication channels which enable a bidirectional transmission of information. Her goal is to learn some global property $a=a(x_1,...,x_N)$ as a function of spatially dispersed information pieces $x_i$. For simplicity we assume that the $N$ locations containing the local information are not mutually connected by any communication channel, i.e. the information cannot flow in between the different regions. This restriction can be understood as one forcing the pieces of information to be truly isolated/localized and removing the dependence on the geometry of the problem. Moreover, we will assume that all the available information carriers (i.e.  physical systems on which information can be encoded) in the scenario are localized at Alice's location. This means that the only way for the information pieces to be transmitted from the $N$ locations towards Alice is by Alice first sending her information carriers towards the locations and then waiting for the carriers (on which information has potentially been encoded) to come back to her.

Assuming that Alice possesses $N$ information carriers, one way for her to learn $a(x_1,...,x_N)$ is thus to send her $N$ signals towards the $N$ locations which will encode the information pieces $x_i$, retrieve them and calculate the function $a$. We denote the time she needs to complete this action (acquisition and transmission of information pieces) as $\tau$ and call it unit time. Suppose now that Alice has limited resources to complete the task, i.e. she has power to send at most $k$ signals per unit time, where $k<N$. This restriction can come from some natural assumption, such as the limited power of the source of information carriers. Upon receiving back the signals, she decodes them and produces an outcome $a$. In general, the information pieces are randomly sampled from some distribution; the process is thus mathematically fully characterized by the following set of conditional probabilities, or $\textit{behavior}$
\begin{equation}\label{eq:Behavior}
\left\{	P\left( a|x_1,...,x_N \right); \quad \forall a\in O; x_1,..,x_N \in I\right\},
\end{equation}
where $I$ and $O$ denote the $\textit{input}$ and $\textit{output}$ alphabets pertaining respectively to $\left\{x_1,...,x_N \right\}$ and to $a$. $P\left( a|x_1,...,x_N \right)$ thus denotes the probability that Alice produces the output $a$ conditioned on the dispersed information being $\left\{x_1,...,x_N \right\}$.\\ 
For a moment, we shall forget that A is using $k$ classical signals, and define the more general notion of $k-way$ $signaling$, which we introduce in a device-independent manner in what follows.
\theoremstyle{definition}
\begin{definition}
A behavior $\left\{ P\left( a|x_1,...,x_N \right), \forall a, x_i\right\}$ is said to be $\textit{k-way signaling}$ $\textit{iff}$ there exists a set of weights $\left\{q_{j_1,...,j_k}, \forall j_1,...,j_k\right\}$ and a set of probability distributions $\left\{P\left( a|x_{j_1},...,x_{j_k} \right), \forall j_1,...,j_k \right\}$ such that the following is satisfied:
\begin{gather*}
P\left( a|x_1,...,x_N \right)=\sum_{j_1,...,j_k} q_{j_1,...,j_k} P\left( a|x_{j_1},...,x_{j_k} \right);\\
\sum_{j_1,...,j_k} q_{j_1,...,j_k}=1;\quad q_{j_1,...,j_k} \geq 0,\quad \forall j_1,...,j_k, 
\end{gather*}
where the domain of the indices $\left\{j_1,...,j_k\right\}$ ranges over all $N \choose k$ subsets of the $N$ locations.
\end{definition}
The intuition behind the latter definition is the following: if the system exhibits $k$-way signaling, it means that its behavior can be modeled by Alice choosing to communicate with locations pertaining to $\left\{x_{j_1},...,x_{j_k}\right\}$ with probability $q_{j_1,...,j_k}$.\\
For example, for $N=3$, a two-way signaling distribution can be decomposed as
\begin{equation*}
\begin{split}
P\left( a|x_1,x_2,x_3 \right)&=q_{12}P\left( a|x_1,x_2\right)+q_{13}P\left( a|x_1,x_3\right)\\
&+q_{23}P\left( a|x_2,x_3\right);
\end{split}
\end{equation*}
\begin{align}
\sum_{i<j}q_{ij}=1; \quad q_{ij} \geq 0, \forall i<j,
\end{align}
where $q_{ij}$ denotes the probability of Alice communicating with locations pertaining to $x_i$ and $x_j$. The definition above will help us to quantify Alice's performance in a device-independent way, thereby providing a fair comparison between classical and quantum resources.\\

\section{Genuine $N$-way signaling}\label{Genuine $N$-way signaling}
In Appendix \ref{app:A} we provide a simple proof that the set of $k$-way signaling behaviors forms a polytope when embedded in a real vector space. Thus, one can characterize this set via necessary and sufficient conditions in form of facet inequalities. These and similar methods have been applied e.g. in the investigation of Bell's inequalities \cite{device} and in causal modelling \cite{Causal1, Causal2}. Providing the full characterization of a polytope is hard in general and will not be pursued here (especially since we want to make a statement about the scenario involving a general number of parties). Instead, here we will focus on a particular inequality that we obtained as a natural generalization of inequalities computed numerically for low $N$ using the Python package \textit{cdd} \cite{cdd}. Moreover, we focus our attention to the case of binary inputs and outputs, i.e. $a,x_i\in\{0,1\}$.\\
The aforementioned inequality is the following:
\begin{equation}\label{eq:ineq}
B\equiv -P(1|0,0,...,0)+\sum_{i=1}^{N}P\left(1|0,...,x_i=1,...0\right)\leq N-1,
\end{equation}
which is satisfied by any $k$-way-signalling behavior, for $k<N$.\\ 
To see that this is the case, notice that any $(N-1)$-way signalling behavior can be expressed as a convex sum of processes which leave out one location from the communication. If the $i$-th location is left out, then
\begin{equation}
P(a|0,0,...,x_i=0,...,0)=P(a|0,0,...,x_i=1,...,0), 
\end{equation} 	
so the first negative term in $B$ cancels at least one of the positive terms and leaves the maximum achievable value equal to $N-1$. The analogous reasoning holds for $k<N-1$. Thus, the violation of this inequality necessarily implies $N$-way signalling. Moreover, notice that even with \textit{no signalling} at all (i.e. $k=0$), Alice can saturate the bound by simply outputting 1. Therefore, the inequality has the particular feature that its bound can be reached without any signalling at all, but cannot be violated unless at least $N$-way-signalling is employed.\\
Note that our inequality represents the probability of successfully accomplishing a task involving information genuinely globally encoded in space. Namely, Alice is supposed to compute the function 
\begin{equation}\label{game}
  a(x_1,...,x_N)=\begin{cases}
    0, & \text{if $x_i=0,\quad \forall i$},\\
    1, & \text{if $\exists j$ s.t. $x_j=1$ and $x_i=0, \forall i\neq j$ },
  \end{cases}
\end{equation}
where the $N+1$ settings $\left\{(0,...,0), (0,...,x_j=1,...,0), \forall j \right\}$ are uniformly distributed. The probability of Alice successfully accomplishing the latter task (winning the game) is  given by the winning probability $P_W$ as follows:
\begin{equation}\label{game1}
P_W\equiv\frac{1}{N+1}\left(P\left(0|0,...,0\right)+\sum_{i=1}^{N}P\left(1|0,...,x_i=1,...0\right)\right).
\end{equation}
Inequality \eqref{eq:ineq} implies that the winning probability is bounded as $P_W \leq P_{bound}=\frac{N}{N+1}$, unless Alice's performance exhibits $N$-way-signalling.\\
Suppose now that Alice possesses limited resources and sends $k<N$ classical signals per unit time $\tau$,  thus achieving $k$-way signalling. In this case, it is clear that she can achieve at best $P_{bound}$ in a single shot experiment. In order to surpass this threshold she needs to send her $k$ signals at least $\lc \frac{N}{k} \rc$ times before producing an output, thereby $\textit{effectively}$ exhibiting $N$-way- signalling behavior.\\ 
\section{Quantum enhancement}\label{Quantum enhanced information speed}
\subsection{Single query}
\begin{figure}[h!]
\centering
\includegraphics[width=\columnwidth]{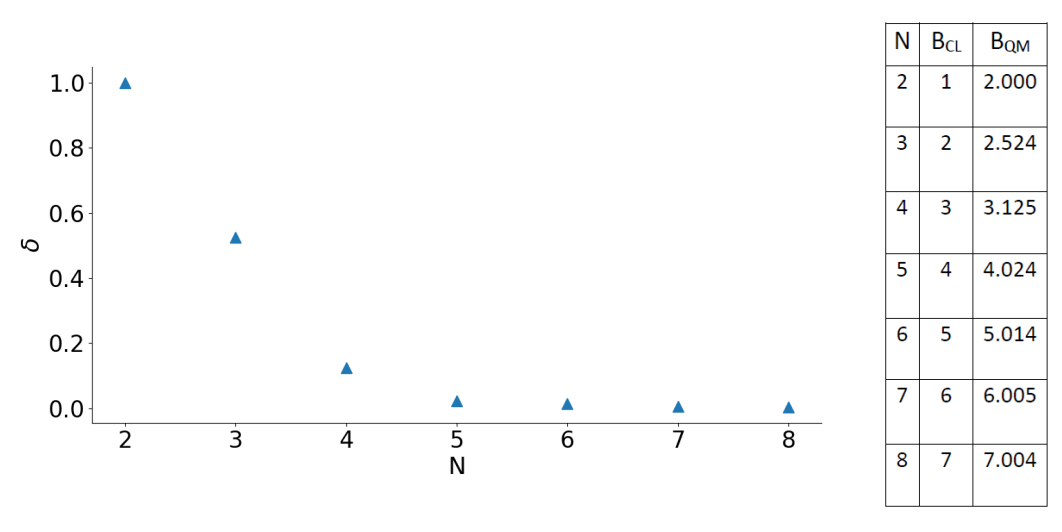}
\caption{ \textbf{Violation of the inequality}. The left graph represents the quantum violation of inequality \eqref{eq:ineq} as function of the number of information locations $N$. The table on the right compares the value of the inequality achievable using classical particles ($B_{CL}$) and the one achievable with a single quantum particle in spatial superposition ($B_{QM}$).}
\label{fig:Fig2}
\end{figure}

In the following we show the possibility of achieving the violation of inequality \eqref{eq:ineq} for arbitrary $N$ by using a single quantum particle (one signal per unit time) prepared in spatial superposition. Here we will focus on the scenario involving a single shot (i.e. single query), while the multiple query case will be addressed later.\\
We assume a simple model where at each location $x_i$ the information is encoded into a box that applies a local phase shift $e^{i x_i \phi_i}$ to the state of the particle, where $\left\{\phi_i\right\}$ are fixed angles known to Alice. The protocol can be summed up as follows. Alice prepares her signal in a uniform spatial superposition of trajectories directed towards the $N$ locations; after interacting with the boxes, the wave packets are bounced back to Alice who performs a binary measurement thereby producing an output $a$.\\
The initial wave function is:\\
\begin{equation}
\ket{\psi_0}=\frac{1}{\sqrt{N}} \sum_n \ket{n},
\end{equation} 
where $\left\{\ket{n}\right\}$ is the basis of spatial modes corresponding to the $N$ trajectories directed towards their pertaining locations.\\
After encoding, the wave function is transformed to
\begin{equation}
\ket{\psi}_{x_1,...,x_N}=\frac{1}{\sqrt{N}} \sum_n e^{i\phi_n x_n} \ket{n}.
\end{equation}
Upon getting back her signals, Alice performs a binary measurement defined by a general POVM $\Pi \equiv \left\{\Pi_0,\Pi_1\right\}$, thereby producing an outcome $a\in \left\{0,1\right\}$.\\
Let us denote the quantum state that arises via encoding when $\left\{x_1=0,x_2=0,...,x_N=0\right\}$ with $\rho_0$, and the one that arises from encoding when $\left\{x_1=0,x_2=0,...,x_i=1,...,x_N=0\right\}$ with $\rho^{(i)}$. Then, if we introduce the following averaged state
\begin{equation}
\rho_1=\frac{1}{N}\sum_{i=1}^{N} \rho^{(i)},
\end{equation}
the left hand side of inequality \eqref{eq:ineq} can be rewritten as
\begin{equation}
\begin{split}
B&=-1+(N+1)\left[ \frac{1}{N+1}\Tr(\Pi_0 \rho_0) + \frac{N}{N+1}\Tr(\Pi_1 \rho_1)    \right]\\
&\equiv -1+(N+1)P_W.
\end{split}
\end{equation}
The expression $P_W=\left[ \frac{1}{N+1}\Tr(\Pi_0 \rho_0) + \frac{N}{N+1}\Tr(\Pi_1 \rho_1)    \right]$ is the probability of successfully distinguishing the quantum states $\rho_0$ and $\rho_1$ given their respective prior probabilities $p_0=\frac{1}{N+1}$ and $p_1=\frac{N}{N+1}$. It is known \cite{Discrimination} that this probability is bounded by 
\begin{equation}
\max\limits_{\Pi}P_W=\frac{1}{2}(1+||p_1\rho_1-p_0\rho_0||_1),
\end{equation}
where $||A||_1$ denotes the trace norm of $A$.\\
The maximum achievable value of $B$ for a given encoding scheme is then given by
\begin{equation}
\max\limits_{\Pi}B=-1+\frac{N+1}{2}(1+||p_1\rho_1-p_0\rho_0||_1)\equiv N-1+\delta,
\end{equation}
where 
\begin{equation}\label{eq:max_bound}
\delta=\frac{1}{2}-\frac{N}{2}+\frac{N+1}{2}||p_1\rho_1-p_0\rho_0||_1
\end{equation} 
is the amount of violation.\\
Let us first analyse the case $N=2$. We set the two boxes' phases to $\phi_{1,2}=\pi$. Alice is supposed to output $a=0$ if both settings are equal to 0 and $a=1$ if one of the settings is equal to 1. Clearly, states $\rho_0$ and $\rho_1$ are mutually orthogonal and thus perfectly distinguishable, thereby enabling Alice to saturate the logical bound of our inequality, i.e. $\delta=1$ (the details are presented in Appendix \ref{app:B}). In contrast, if she uses one classical signal per unit time, she needs double the time to achieve a violation. Therefore, spatial coherence doubles the information acquisition speed involved in completing the task.\\ 
The $N\geq3$ case is more complicated and the detailed analysis is presented in Appendix \ref{app:C}. In order to analytically demonstrate the possibility of violating the inequality, we set $\lc \frac{N}{2} \rc$ of the boxes' phases to some angle $\phi$ and the rest to $-\phi$; we show that a clear violation $\delta>0$ is achieved for $\cos(\phi)>\frac{N(N-6)+4}{(N-2)^2}$ (with a small correction for odd $N$). The numerical results of the violation are shown in Figure \ref{fig:Fig2}.


We have therefore proved the possibility of genuine multi-way signalling with an arbitrary number of locations using one particle in spatial superposition within one unit of time $\tau$. On the other hand, suppose that Alice possesses a source producing one particle with a defined trajectory per unit time: in this case, she cannot achieve the quantum performance for this task even in $\left(N-1\right)\tau$ time, since she can communicate with maximally one location per unit time. Hence, spatial coherence as a resource provides an arbitrarily large enhancement of the information acquisition speed as defined by our task and inequality \eqref{eq:ineq}. Note that there is no conflict with special relativity, since the information carriers' speed is limited by the speed of light, which is reflected in the time $\tau$ that light needs to travel back and forth from Alice to the boxes.\\ 
\subsection{Multiple queries} 
In the previous subsection we provided a proof of the possibility of achieving arbitrarily high levels of signalling using a single quantum particle in a single shot. However, the violation of the inequality scales poorly for large $N$, as seen in Fig. \ref{fig:Fig2}. In what follows, we will analyse how Alice's performance in the quantum case improves by allowing her to send her particle multiple times towards the parties, i.e. by relaxing the restriction of one unit of time $\tau$. In order to show a clear gap between the classical and the quantum performance, we will analyse a variation of the task/game \eqref{game1}: namely, the settings and goals of the game will remain the same, but we will modify the prior probabilities of the settings. More precisely, in the task/game \eqref{game} the $(N+1)$ settings $\left\{(0,...,0), (0,...,x_j=1,...,0), \forall j \right\}$ were uniformly distributed, i.e. each of the settings was sampled with probability $\frac{1}{N+1}$; on the other hand, in this section we will modify the priors as follows: the setting $(0,...,0)$ will be sampled with probability $\frac{1}{2}$, and each of the $N$ remaining settings $\left\{(0,...,x_j=1,...,0);\forall j) \right\}$ will be sampled with probability $\frac{1}{2N}$. The winning probability for the new task is thus given by:
\begin{equation}\label{game2}
P_W=\frac{1}{2}P\left(0|0,...,0\right)+\frac{1}{2}\frac{1}{N}\sum_{i=1}^{N}P\left(1|0,...,x_i=1,...0\right).
\end{equation}
Now, if Alice possesses one classical particle/signal and has at disposal $k$ units of time $\tau$ for communication, then she can exhibit at most $k$-way signalling. For instance, if she communicates with the first $k$ parties, then the probability distributions obey the following:
\begin{equation}
\begin{split}
P\left(a|x_1,...,x_k,x_{k+1},...x_{N}\right)&=P\left(a|x_1,...,x_k,x_{k+1}',...,x_{N}'\right),\\
&\forall a,x_i,x_j'=0,1.
\end{split}
\end{equation}
The best strategy Alice can employ is to output 0 if $\left\{x_i=0, i=1,...,k\right\}$ and to output 1 otherwise. The maximum winning probability after $k$ classical queries is thus:
\begin{equation}\label{win_game_1}
P_{Cl}(k)=\frac{1}{2}\left(1+\frac{k}{N}\right).
\end{equation}
Next, suppose that Alice can prepare her particle in spatial superposition and query it through the boxes $k$ times (in $k$ units of time $\tau$). After each query, she can apply a fixed unitary transformation, and at the end of the $k$ queries she performs a measurement. We will show that Grover's algorithm (with a modified final measurement) can give her the same speed-up for the execution of her task as the well known $O(\sqrt{N})$ speed-up obtained in Grover's search \cite{Grover1, Grover2}.\\
Let us suppose that the $N$ locations/boxes encode their local bits via $\pi$-phase-shifts $\phi_i=x_i\pi$ and that the initial state prepared by Alice is a uniform superposition $\ket{\psi_0}=\frac{1}{\sqrt{N}} \sum_n \ket{n}$. Moreover, after each query, Alice applies the following unitary transformation (also known as the \textit{inversion about mean operation} \cite{Grover2}):
\begin{equation}
U=2\ket{\psi_0}\bra{\psi_0}-\mathbbm{1}.
\end{equation}
If all the settings are 0, then the final state after $k$ runs is $\ket{\psi_0}$, since $U\ket{\psi_0}=\ket{\psi_0}$. On the other hand, if the $i$-th setting is $x_i=1$ and all the others are 0, then the final state after $k$ queries is equal to the one obtained after $k$ queries of Grover's algorithm \cite{Grover2}:
\begin{equation}
\ket{\psi_k^{(i)}}=\cos\left(\frac{2k+1}{2}\theta\right)\ket{\bar{i}}+\sin\left(\frac{2k+1}{2}\theta\right)\ket{i},
\end{equation}
where $\ket{\bar{i}}\equiv\frac{1}{\sqrt{N-1}}\sum_{j\neq i}\ket{j}$ and $\sin\left(\theta/2\right)=1/\sqrt{N}$. Setting $k=\frac{\pi}{4}\sqrt{N}$ and taking the limit of large $N$ one obtains
\begin{equation}
\ket{\psi_k^{(i)}}\approx \ket{i}-\frac{1}{\sqrt{N}}\ket{\bar{i}}.
\end{equation}
Therefore, the winning probability \eqref{game2} after $O(\sqrt{N})$ queries is
\begin{equation}\label{value}
P_W=\frac{1}{2}\left(\Tr(\Pi_0 \rho_0)+\Tr(\Pi_1 \rho_1) \right),
\end{equation}
where $\left\{\Pi_0,\Pi_1\right\}$ is a POVM, $\rho_0=\ket{\psi_0}\bra{\psi_0}$, and 
\begin{equation}
\rho_1= \frac{1}{N}\sum_i \ket{i}\bra{i}+O(N^{-3/2})=\frac{1}{N}\mathbbm{1}+O(N^{-3/2}).
\end{equation} 
The maximum achievable value of expression \eqref{value} is determined by the Helstrom bound:
\begin{equation}
\max\limits_{\Pi}P_W=\frac{1}{2}\left(1+\frac{1}{2}||\rho_1-\rho_0||_1\right).
\end{equation}
Using the fact that the trace norm of an operator is the sum of the absolute values of its eigenvalues we obtain
\begin{equation}
\max\limits_{\Pi}P_W=1-\frac{1}{2N}+O(N^{-3/2}).
\end{equation}
Therefore, Alice needs only $O(\sqrt{N})$ queries to achieve the task with unit probability (for large $N$), which thereby provides a $O(\sqrt{N})$ speed-up with respect to the classical case; in other words, spatial coherence increases the information acquisition speed by $O(\sqrt{N})$. Note that we did not prove the optimality of the latter protocol; it might be the case that one can achieve a better speed-up with a different encoding of local bits and a different intermediate unitary transformation $U$.\\

\section{Discussion}
We started the previous section by proving the possibility of exhibiting $N$-way-signalling with a single quantum particle in superposition by violating an inequality which separates $(N-1)$ from $N$-way-signalling behaviors. Since the violation of the inequality drops quickly to 0 for large $N$, this violation ought to be seen as a proof of principle, rather than as a compelling application. Moreover, the $(N-1)$-way signalling bound in \eqref{eq:ineq} can be saturated already with a random guess (i.e. one does not gain any advantage by using $(N-1)$ classical signals with respect to no signals at all). As a response to the latter drawback, we introduced the game \eqref{game2}, where we chose different prior probabilities with respect to the preceding game in order to obtain a scaling of the probability of success with the amount of signalling. Indeed, the probability of success \eqref{win_game_1} scales linearly with the amount of signalling, i.e. with the number of classical queries. We then proceeded by showing that a Grover-like algorithm provides a quadratic enhancement of the information acquisition speed.\\
The latter can be understood from a more practical point of view as an oracle problem that can be solved more efficiently using quantum resources. Here we want to emphasize the method we came up with this new oracle problem and its pertaining quantum speed-up. We started off with the definition of $k$-way-signalling behaviors and showed that they constitute a polytope and can consequently be characterized in terms of facet inequalities. Generally, each inequality corresponds to a particular game and can therefore be interpreted as an oracle problem. This generic method can perhaps be used to discover new oracle problems with a potential quantum advantage by defining a polytope structure of conditional probability distributions $P\left(a|x_1,...,x_N\right)$ (e.g. in our particular case: the set of $k$-way-signalling distributions) and investigating its facet inequalities. The latter inequalities can be interpreted as oracle problems, for which one can try to obtain a quantum advantage. The converse research direction is also interesting to pursue, i.e., can every oracle problem (with suitably chosen prior probabilities) be seen as a facet inequality of a certain polytope? These connections can prove useful in shedding light on the structure of oracle problems in general, and can provide further intuition on the quantum mechanical speed-up obtained in these problems.\\
An interesting remark about our scheme is the fact that the only resource used is coherence of paths and all the information is encoded via local phases. It still remains open whether internal degrees of freedom could increase the performance. Moreover, in this work we have not analysed the scenario in which Alice sends more quantum signals/particles per unit time, in the case of which one might expect further advantage arising from the entanglement between the particles. We leave these considerations for future work.\\

\renewcommand\refname{Bibliography}


\begin{acknowledgements}
BD acknowledges support from an ESQ Discovery Grant of the Austrian Academy of Sciences (OAW) and the Austrian Science Fund (FWF) through BeyondC-F7112. SH acknowledges support from the Erasmus Programme. Both authors thank Ämin Baumeler, Joshua Morris and Francesco Massa for helpful discussions.
\end{acknowledgements}

\appendix

\section{$K$-way signaling behaviors form a polytope}\label{app:A}
A behavior $\left\{	P\left( a|x_1,x_2,...,x_N \right); \forall a\in O; x_1,..,x_N \in I\right\}$ can be regarded as an element of a real vector space of dimension $D=LK^N$, where $K$ and $L$ are cardinalities of the input and output alphabets. In what follows we prove that the subset $S_k$ of all $k$-way signaling behaviors forms a polytope.\\
A general conditional probability can be written as a convex sum of deterministic distributions
\begin{equation}
P(a|b)=\sum_f \mu_f \delta_{a,f(b)}
\end{equation}
where the sum runs over all functions $\textit{f}: I \rightarrow O$, where $I$ and $O$ are respectively the input and output alphabets to which $\textit{b}$ and $\textit{a}$ pertain. A general $k$-way signaling correlation can thus be expressed as
\begin{equation}
\begin{split}
&P\left( a|x_1,x_2,...,x_N \right)\\
&= \sum_{j_1,j_2,...,j_k} q_{j_1,j_2,...,j_k} P\left( a|x_{j_1},x_{j_2},...,x_{j_k} \right)\\
&=\sum_{j_1,j_2,...,j_k} q_{j_1,j_2,...,j_k} \sum_f \mu_{f_{j_1,...,j_k}} \delta_{a,f_{j_1,...,j_k}(x_{j_1},...,x_{j_k})}\\
&=\sum_{f,j_1,...,j_k} \lambda_{f,j_1,...,j_k} \delta_{a,f_{j_1,...,j_k}(x_{j_1},...,x_{j_k})},
\end{split}
\end{equation}
where we defined a new set of weights
\begin{equation}
\begin{split}
&\lambda_{f,j_1,...,j_k} \equiv \mu_{f_{j_1,...,j_k}} q_{j_1,j_2,...,j_k} ; \quad \lambda_{f,j_1,...,j_k} \geq 0,\\
&\sum_{f,j_1,...,j_k} \lambda_{f,j_1,...,j_k} =1. 
\end{split}
\end{equation}
Therefore, any $k$-way signaling behavior can be written as a convex combination of a finite number of deterministic behaviors; moreover, any convex combination of $k$-way signalling behaviours is another $k$-way signalling behaviour. The latter statements imply that $S_k$ is a polytope.\\

\section{Quantum violation for $N=2$}\label{app:B}
Suppose we have only two boxes which provide phase shifts $e^{i \phi_i x_i}$ depending on their local bits $x_i$ and let both phases be set to $\pi$.\\
Alice sends her photon in a homogeneous superposition
\begin{equation}
\ket{\psi_0}=\frac{1}{\sqrt{2}} \left( \ket{1} + \ket{2} \right),
\end{equation} 
where $\ket{1/2}$ represent states of defined trajectories directed towards boxes 1 and 2 respectively. Since the boxes' phases are fixed to $\pi$, the state Alice receives after the interaction with the boxes is 
\begin{equation}
\ket{\psi_{x_1,x_2}}=\frac{1}{\sqrt{2}} \left[ (-1)^{x_1}\ket{1} + (-1)^{x_2}\ket{2} \right].
\end{equation} 
Plugging $\rho_{x_1x_2}=\ket{\psi_{x_1,x_2}}\bra{\psi_{x_1,x_2}}$ into expression \eqref{eq:max_bound}, we obtain $\max B=2$, which saturates the logical bound of the inequality. The latter follows directly from the orthogonality of states $\ket{\psi_{00}}$ and $\ket{\psi_{01/10}}$ which enables perfect state discrimination. Specifically, the required measurement is given by a projection on vectors $\ket{\pm}=\frac{1}{\sqrt{2}} \left[ \ket{1} \pm \ket{2} \right]$.\\

\section{Proof of the quantum violation for arbitrary $N$}\label{app:C}
In this appendix we provide a step by step proof of the violation of inequality \eqref{eq:ineq} for $N\geq2$ by using a single particle. According to the encoding scheme portrayed in section $\textit{Quantum enhancement}$, the density operators of interest are
\begin{equation}
\rho_0=\frac{1}{N}\sum_{n,m}\ket{n}\bra{m}
\end{equation}
and
\begin{equation}
\begin{split}
\rho_1&=\frac{1}{N^2}\sum_{n,m}\ket{n}\bra{m}\sum_k e^{i(\phi_n\delta_{n,k}-\phi_m\delta_{m,k})}\\
&=\frac{1}{N^2}\sum_{n,m}\ket{n}\bra{m} \left[N+(1-\delta_{n,m})(e^{i\phi_n}+e^{-i\phi_{m}}-2)\right].
\end{split}
\end{equation}
The goal is to calculate the trace norm of the following operator
\begin{equation}\label{eq:operator}
\begin{split}
p_1\rho_1-p_0\rho_0&=\frac{1}{N(N+1)}\sum_{n,m}\ket{n}\bra{m}[ (N-3) + e^{i\phi_n}+e^{-i\phi_m}\\
&+ 2\delta_{n,m}-\delta_{n,m}\left( e^{i\phi_n}+e^{-i\phi_m} \right) 	].
\end{split}
\end{equation}
The trace norm of an operator $M$ can be expressed succinctly as 
\begin{equation}
||M||_1 = \sum_i |\lambda_i|,
\end{equation}
where $\lambda_i$ are the eigenvalues of the given operator. The calculation has thus been reduced to an eigenvalue problem.\\
Let's further specify the encoded phases by setting half of them equal to an arbitrary phase $\phi$ and the other half to $-\phi$. More specifically, if $N=2K$ for some $K \in \mathbb{N} $, set $K$ of them to $\phi$ and $K$ of them to $-\phi$, while if $N=2K+1$, set $K+1$ of them to $\phi$ and $K$ of them to $-\phi$. The operator \eqref{eq:operator} is then equal to
\begin{equation}\label{eq:operator2}
\begin{split}
p_1\rho_1-p_0\rho_0&=\frac{1}{N+1}[\frac{2}{N}(1-\cos(\phi))\mathbbm{1}\\
&+(N-3)\ket{\psi_0}\bra{\psi_0}+\ket{\phi}\bra{\psi_0}+\ket{\psi_0}\bra{\phi}   ],
\end{split}
\end{equation}
where we introduced an auxiliary $\textit{phase vector}$  
\begin{equation}
\ket{\phi}\equiv \frac{1}{\sqrt{N}}\sum_n e^{i\phi_n}\ket{n}.
\end{equation}
Let's define 
\begin{equation}
M\equiv (N-3)\ket{\psi_0}\bra{\psi_0}+\ket{\phi}\bra{\psi_0}+\ket{\psi_0}\bra{\phi}   
\end{equation}
and diagonalize it in the two-dimensional subspace. Two orthogonal vectors in the given subspace are
\begin{equation}
\begin{split}
&\ket{v}\equiv\ket{\psi_0},\\
&\ket{w}\equiv   \frac{1}{\sqrt{1-|\braket{\psi_0|\phi}|^2}} \left( \ket{\phi} - \braket{\psi_0|\phi} \ket{\psi_0} \right).
\end{split}
\end{equation}
We're going to treat even and odd $N$ cases separately. Starting with even $N=2K$, the following holds:
\begin{equation}
\begin{split}
&\ket{\psi_0}=\ket{v},\\
&\ket{\phi}=\cos(\phi)\ket{v} + \sin(\phi)\ket{w},
\end{split}
\end{equation}
where we used 
\begin{equation}
\braket{\psi_0|\phi}=\frac{1}{N}\frac{N}{2}(e^{i\phi}+e^{-i\phi})=\cos(\phi), 
\end{equation}
and assumed that $\sin(\phi)\geq 0$ (which will be consistent with the end result).\\
Substituting the latter into $M$ we obtain:
\begin{equation}
M =  (N-3+2\cos(\phi))\ket{v}\bra{v}+ \sin(\phi)\ket{v}\bra{w}  + \sin(\phi)\ket{w}\bra{v}.
\end{equation}
The eigenvalues are 
\begin{equation}
\begin{split}
&\lambda_{\pm} =  \frac{1}{2}\left[  A \pm \sqrt{ A^2  +  (2\sin(\phi))^2  }      \right],\\
&A\equiv N-3+2\cos(\phi).
\end{split}
\end{equation}
Now we have to return to the full operator \eqref{eq:operator2}; since the identity matrix is diagonal in any basis, the eigenvalues trivially follow: two of them are equal to 
\begin{equation}
\begin{split}
\mu_{+}=\frac{1}{N+1}\left( \frac{2}{N}(1-\cos{\phi})  + \lambda_{+}    \right),\\
\mu_{-}=\frac{1}{N+1}\left( \frac{2}{N}(1-\cos{\phi})  + \lambda_{-}    \right),
\end{split}
\end{equation}
and $(N-2)$ of them which correspond to eigenvectors orthogonal to our two-dimensional subspace are equal to 
\begin{equation}
\mu_{i}=\frac{1}{N+1} \frac{2}{N}(1-\cos{\phi}).
\end{equation}
The trace norm is then
\begin{equation}
\begin{split}
&||p_1\rho_1-p_0\rho_0||_1=\sum_j |\mu_j|= \\
&= \frac{1}{N+1} ( (N-2)|\frac{2}{N}(1-\cos(\phi))|\\
&+ \abs*{\frac{2}{N}(1-\cos(\phi))+\lambda_{+}}+ \abs*{\frac{2}{N}(1-\cos(\phi))+\lambda_{-}}				).
\end{split}
\end{equation}
If $|\frac{2}{N}(1-\cos(\phi))|>|\lambda_{-}|$ holds, then $\delta$ turns out to be 0 and independent of $\phi$, hence not violating the $(N-1)$-way signaling bound.\\
On the contrary, if $|\frac{2}{N}(1-\cos(\phi))|<|\lambda_{-}|$:
\begin{equation}
\delta=\frac{3}{2}-\frac{N}{2}-\frac{2}{N}+\frac{2}{N}\cos(\phi)-\cos(\phi)+\frac{1}{2}\sqrt{A^2+(2\sin(\phi))^2}.
\end{equation}
Inserting the assumed inequality in the previous expression we get
\begin{equation}
\begin{split}
\delta&>\frac{3}{2}-\frac{N}{2}-\cos(\phi)+\frac{1}{2}\left[	A-\sqrt{A^2+(2\sin(\phi))^2}\right]\\
&+\frac{1}{2}\sqrt{A^2+(2\sin(\phi))^2}=0,
\end{split}
\end{equation}
which means that the inequality is violated. Now it only remains to be shown that for any even $N$ there exists $\phi$ such that $|\frac{2}{N}(1-\cos(\phi))|<|\lambda_{-}|$ is satisfied.\\
Rearranging and squaring the inequality, we obtain 
\begin{equation}
\left[\frac{4}{N}(1-\cos(\phi))+ A	\right]^2 < A^2+(2\sin(\phi))^2.
\end{equation}
A few trigonometric manipulations lead to
\begin{equation}
\frac{8}{N^2}\sin^2\left(\frac{\phi}{2}\right)\left\{ (N-2)^2\cos(\phi)-N(N-6)-4\right\}>0,
\end{equation}
which is satisfied if
\begin{equation}
\cos(\phi)>\frac{N(N-6)+4}{(N-2)^2}.
\end{equation}
This means that for any even $N$ it is possible to find $\phi$ such that our communication scheme violates the $(N-1)$-way signaling bound. In particular, $\phi$ has to be chosen such that $\frac{N(N-6)+4}{(N-2)^2}<\cos(\phi)<1$.\\
The previous analysis holds for even $N$; for odd $N=2K+1$ we get
\begin{equation}
\braket{\psi_0|\phi}=\frac{1}{N}\left( (N-1)\cos(\phi) + e^{i\phi} \right)=\cos(\phi)+ \frac{i}{N} \sin(\phi), 
\end{equation}
and
\begin{equation}
\begin{split}
M' &=  (N-3+2\cos(\phi))\ket{0}\bra{0}+ \sin(\phi)\sqrt{1-\frac{1}{N^2}}\ket{0}\bra{1} \\
& + \sin(\phi)\sqrt{1-\frac{1}{N^2}}\ket{1}\bra{0},
\end{split}
\end{equation}
which is equivalent to the even $N$ case up to the factor $\sqrt{1-\frac{1}{N^2}}$ in the off-diagonal elements. Following the analogous procedure, one obtains a clear violation $\delta>0$ if $|\frac{2}{N}(1-\cos(\phi))|<|\lambda_{-}|$, where $\lambda_{-}$ is the negative eigenvalue of the operator $M'$. The assumed inequality can be cast in a simpler form using trigonometric relations and can be shown to be equivalent to the condition
\begin{equation}
\cos(\phi)>\frac{N(N-6)+5}{N^2-2N+3}.
\end{equation}
Therefore, we showed the possibility of achieving multi-way signaling with an arbitrary number of parties.\\

\end{document}